\documentclass{JHEP3}

\usepackage{url,amsmath,amsthm,amscd,graphicx,psfrag}
\usepackage[all]{xy}

\def\tX{{\tilde X}}
\def\tV{{\tilde V}}
\def\pt{{\rm pt}}
\def\dirac{\slash{\! \! \! \! D}}
\def\ad{{\rm ad}}
\def\bH{{\bar H}}
\def\coker{{\rm coker}}
\def\Ima{{\rm Im}}
\def\Tr{{\rm Tr}}

\let\a=\alpha   \let\b=\beta    \let\g=\gamma   \let\d=\delta
              
   \let\l=\lambda  \let\m=\mu      
\let\n=\nu                       
\let\t=\tau

  \def\co{{\cal O}}

 \def\IC{{\mathbb C}} \def\IP{{\mathbb P}}
  
\def\IZ{{\mathbb Z}}

\theoremstyle{definition}

\theoremstyle{plain}

\theoremstyle{remark}

\newtheorem*{ackn}{Acknowledgments}

\def\plb#1 #2 {Phys. Lett. {\bf B#1} #2 }
\def\phr#1 #2 {Phys. Rep. {\bf  #1} #2 }
\def\npb#1 #2 {Nucl. Phys. {\bf B#1} #2 }
\def\aph#1 #2 {Ann. Phys. {\bf #1} #2 }
\def\jmp#1 #2 {J. Math. Phys. {\bf #1} #2 }
\def\jgp#1 #2 {J. Geom. Phys. {\bf #1} #2 }
\def\prd#1 #2 {Phys. Rev. {\bf D#1} #2 }
\def\prl#1 #2 {Phys. Rev. Lett. {\bf #1} #2 }
\def\rmp#1 #2 {Rev. Mod. Phys.  {\bf #1} #2 }
\def\zpc#1 {Z. Phys. {\bf #1C} }
\def\cmp#1 #2 {Commun. Math. Phys. {\bf #1} #2 }
\def\cqg#1 #2 {Class.Quant.Grav. {\bf #1} #2 }
\def\mpl#1 {Mod. Phys. Lett. {\bf A#1} }
\def\cpc#1 {Computer Phys. Commun. {\bf #1} }
\def\ijmp#1 {Int. J. Mod. Phys. {\bf A#1} }
\def\ijmpC#1 {Int. J. Mod. Phys. {\bf C#1} }
\def\atmp#1 {Adv. Theor. Math. Phys. {\bf #1} }

\numberwithin{equation}{section}

\let\0=\over
\def\2{{1\over2}}
\let\6=\partial
\def\({\left(}       \def\){\right)}

\let\bra=\langle        \let\ket=\rangle        \def\<#1\>{\bra #1 \ket}
   
\preprint{UPR-1146-T}
\title{Tri-linear Couplings in an Heterotic Minimal Supersymmetric Standard Model}

\author{Vincent Bouchard\\
The Mathematical Sciences Research Institute\\
17 Gauss Way\\
Berkeley, CA 94720-5070\\
E-mail: \email{vincentb@msri.org}}

\author{Mirjam Cveti\v c\\
Department of Physics and Astronomy\\
University of Pennsylvania\\
Philadelphia, PA 19104-6396\\
E-mail: \email{cvetic@physics.upenn.edu}}

\author{Ron Donagi\\
Department of Mathematics\\
University of Pennsylvania\\
Philadelphia, PA 19104-6395\\
E-mail: \email{donagi@math.upenn.edu}}

\abstract{We  calculate, at the classical level, the superpotential  tri-linear
couplings of the only known globally consistent heterotic minimal supersymmetric Standard Model \cite{Bouchard:2005ag}. This recently constructed model is based on a compactification of the $E_8\times E_8$ heterotic string theory on
a Calabi-Yau threefold with $\IZ_2$ fundamental group, coupled with a
slope-stable holomorphic $SU(5)$ vector bundle. In the observable sector the
massless particle content is that of the three-family supersymmetric Standard
Model with $n=0,1,2$ massless Higgs pairs, depending on  the location in the
vector bundle moduli space, and no exotic particles. We obtain non-zero Yukawa
couplings for the three up-sector quarks, and vanishing R-parity violating
terms. In particular, the proton is stable. Another interesting feature is the existence of tri-linear couplings, on
the loci with  massless Higgs pairs, generating $\mu$-mass
parameters for the Higgs pairs and neutrino
mass terms, with specific vector bundle moduli playing  the role of
right-handed neutrinos.}

\begin{document}

\section{Introduction}
Consistent four-dimensional solutions of string theory play an important role
as a link between string theory and  particle physics. One particular goal of
this program  is to  derive  consistent supersymmetric solutions of string
theory with the massless particle content of the three-family Standard Model.
On the heterotic string theory side, compactifications of the
 $E_8\times  E_8$  heterotic string theory on Calabi-Yau
threefolds with stable holomorphic vector
 bundles
 yielded large classes of supersymmetric
 three-family standard-like models (see \cite{Donagi:2004ub} and references therein). For the minimal supersymmetric standard-like model constructions with 
heterotic conformal field theory orbifolds, see \cite{CleaverBuchmuller}.
 On the other hand, within the framework of Type II string theory,
compactifications with intersecting D-branes on toroidal orbifolds also
provide  a wealth of three-family supersymmetric standard-like model constructions
where non-Abelian gauge symmetry, chirality and family replication
have a geometric origin (see \cite{CSUrev} for a review and
references therein). For standard model-like constructions  with rational 
conformal field theory orientifolds, see \cite{Dijkstra:2004cc} and 
references therein.

 In spite of significant progress  made in developing the techniques for these constructions and a
 proliferation of models with semi-realistic particle physics features,  most of the
 constructions  suffer from some phenomenological deficiencies.
 In addition to the  minimal supersymmetric Standard Model (MSSM) particle  content, the models typically
  possess more than one Higgs  doublet pair  and additional  Standard Model
  chiral exotics.
  Furthermore,  consistent constructions with the supersymmetric Standard Model
 particle content {\it and } stabilized moduli remain elusive, though there has been
 progress  made on the Type IIB side (see e.g. \cite{CSUrev} and references
 therein).

 Another important  test of  these constructions is at the level of couplings.
  In particular,  the  tri-linear  couplings  of the matter chiral superfields
 test the string theory predictions for   Yukawa couplings of the Standard Model.
 For the heterotic string theory compactifications on Calabi-Yau threefolds
such couplings can be calculated, in the classical limit,  by determining the
non-zero triple pairings of the cohomology group elements which determine the
massless particle spectrum.   There has also been progress made in the
calculation of  quantum contributions to such couplings, i.e. non-perturbative
worldsheet instanton contributions,  which involve the pairing of quantum
cohomology groups (see e.g. \cite{sharpe} and references therein). Within this
framework  there remains  an outstanding problem of determining the moduli
dependence of the K\"ahler potential for the matter chiral superfields, which
in turn  determines the normalization of the kinetic energy terms for the
matter fields. On the Type IIA side, for toroidal orbifold compactifications
with
   intersecting D6-branes,  conformal field theory techniques allow for the full tree-level calculations of such
   couplings, including the superpotential and K\"ahler potential contributions to Yukawa couplings  (see
    \cite{CPetal} and references therein).

Recently, major progress has been achieved by constructing a specific, globally
consistent supersymmetric  solution  of  heterotic string theory, that yields
a massless spectrum with  minimal supersymmetric Standard Model particle
content and no exotic particles \cite{Bouchard:2005ag}. The construction is
obtained by compactifying the $E_8\times E_8$ heterotic string theory on a
Calabi-Yau threefold  $X$ with $\IZ_2$ fundamental group, coupled with a
stable holomorphic $SU(5)$ vector bundle $V$. In the observable sector the
massless particle content is that of the three-family supersymmetric Standard
Model (MSSM) with $n=0,1,2$ massless Higgs pairs. The value of $n$ depends on
the location in the vector bundle moduli space.   The model also possesses a
number of K\"ahler and complex structure moduli of the Calabi-Yau threefold,
and a number of vector bundle moduli.

The Green-Schwarz anomaly cancellation (global consistency) condition requires
that the difference of the second Chern classes $c_2(TX)-c_2(V)-c_2(U)=[W]$,
where $V$ is the visible bundle and $U$ the hidden bundle, be the effective
class $[W]$ of an holomorphic curve, around which $M5$-branes are wrapped. In
the construction of \cite{Bouchard:2005ag} the hidden bundle is chosen to be
the trivial bundle,  which implies that the hidden sector has the
 spectrum of  $N=1$ $E_8$ super Yang-Mills theory. In this case,
 the anomaly condition requires that $c_2(TX)-c_2(V)=[W]$ be effective,
 which is indeed satisfied in this particular construction. In conclusion,
 the model of \cite{Bouchard:2005ag} is a manifestly
supersymmetric, globally consistent solution of heterotic string theory.

 Another heterotic string theory construction \cite{Braun:2005nv}
 that yields the massless spectrum of the MSSM (with an extra $U(1)$ factor),
 based on a slope-stable visible vector bundle $V$ \cite{Braun:2006ae}, has been claimed. However, as it stands, this construction does not satisfy the Green-Schwarz
anomaly cancellation  condition, and thus it is {\it not} globally consistent.
More precisely, in this model $c_2 (TX) - c_2 (V)= [W]$ is not an effective
class. Therefore, the hidden bundle $U$ cannot be trivial, and one must find a
slope-stable hidden vector bundle $U$ such that $c_2(TX)-c_2(V)-c_2(U)=[W]$ is
effective.
 Currently, there are no known examples of such slope-stable vector bundles;
 it was shown in \cite{Gomez:2005ii} that the proposed hidden bundles $U$ that
 satisfy the Green-Schwarz anomaly cancellation condition \cite{Braun:2005ux,Braun:2005nv}, are not slope-stable.
  It was further suggested in \cite{Braun:2006ae}
  that the introduction of anti-$M5$ branes may cancel the Green-Schwarz anomaly.
However, in this context the introduction of  anti-$M5$ branes should render the solution physically unstable, leading to annihilation  processes with $M5$-branes and/or vector bundle states.

The main purpose of this paper is to confront the predictions of the globally
consistent model of \cite{Bouchard:2005ag} at the level of  tri-linear
superpotential couplings for the chiral matter superfields and to  study the
implications of these couplings for particle phenomenology. The calculation of
the tri-linear superpotential couplings for the matter superfields is
performed in the classical limit. The model has non-zero Yukawa couplings for
all three up-sector quarks, and depending on the location in the moduli space
a suitable mass hierarchy for the up-sector quarks may be achieved. At the classical level the couplings to the
down-quarks are zero,  though the expectation
 is that such couplings would be non-zero at the quantum level.
 Moreover, all R-parity violating couplings vanish at tree level. In particular, baryon number and lepton number violating processes that could lead to physically unacceptable rapid proton decay are absent; thus, at tree level the proton is stable (see \cite{Nath:2006ut}). The requirement that R-parity violating couplings also vanish at the quantum level should further
 constrain  phenomenologically viable solutions by restricting the allowed
subspace of vector bundle moduli.

Another interesting feature of these model is the existence of tri-linear
couplings of the  up-Higgs fields,  down-Higgs or lepton doublet fields,  and
vector bundle  moduli. We calculate  these couplings on the loci with
$n=1$ and $n=2$ massless Higgs pairs.  Our results
 demonstrate that  specific directions in moduli space,
 perpendicular to the locus with $n$ massless Higgs  pairs, generate $\mu$-terms for the Higgs pairs.
  In the effective field theory, this is demonstrated by giving a non-zero vacuum
 expectation value to a  specific linear combination of vector bundle moduli
 which in some tri-linear couplings generates the $\mu$ parameters  for the $n$
Higgs pairs. There are also additional couplings of the down-Higgs fields to one
lepton doublet and specific vector bundle moduli, which  can be interpreted as right-handed neutrinos.
These couplings in turn provide mass terms for neutrinos.

The paper is organized as follows. In section \ref{s:constr} we summarize the
construction of \cite{Bouchard:2005ag}. In section \ref{s:lep} we briefly
review the features of the MSSM,  and  the low-energy physics obtained in
the minimal supersymmetric Standard Model of the heterotic string.  Section
\ref{s:moduli} is devoted to the computation of the vector bundle moduli of
the model. This section provides important prerequisite results for the
superpotential coupling calculations. In section \ref{s:comp} we compute the
tri-linear superpotential couplings.  We give a phenomenological
interpretation of the effective theory couplings in section \ref{s:pheno}.

\section{Construction}\label{s:constr}

In this section we briefly review the $SU(5)$ heterotic standard model introduced in \cite{Bouchard:2005ag}. More details can be found in \cite{Bouchard:2005ag}, and in previous papers \cite{Donagi:2000si,Donagi:2000sm,Donagi:2000zf,Donagi:2004ub} where different bundles on the same manifold were constructed.

In this model we compactify the $E_8 \times E_8$ heterotic string
theory on a Calabi-Yau threefold $X$ with $\IZ_2$ fundamental group.
 Moreover, we construct a stable $SU(5)$ bundle on $X$ which is twisted by a $\IZ_2$
 Wilson line to break the visible $E_8$ gauge group to the standard model gauge group
  $SU(3)_C \times SU(2)_L \times U(1)_Y$.

\subsection{The Manifold}

The non-simply connected Calabi-Yau threefold $X$ is constructed by considering a simply connected Calabi-Yau threefold $\tX$, elliptically fibered over a rational elliptic surface $B$, that admits a free $F=\IZ_2$ action preserving the fibration. The quotient $X=\tX/F$ is a Calabi-Yau threefold, has fundamental group $F$ and is a genus-one fibration.

Let $B'$ be a rational elliptic surface, and let $\tilde X$ be a Calabi-Yau threefold with an elliptic fibration $\pi: {\tilde X} \to B'$ (we also require that $\pi$ has a section). The manifold $\tilde X$ also admits a description as a fiber product $B \times_{\IP^1} B'$ of two rational elliptic surfaces $B$ and $B'$ over $\IP^1$:
\begin{equation}
{\tilde X} = \{ (p,p') \in B \times B' | \b'(p') = \b (p) \},
\end{equation}
 where $\b: B \to \IP^1$ and $\b': B' \to \IP^1$ are the elliptic fibrations of the rational elliptic surfaces $B$ and $B'$.

Thus $\tilde X$ can be described by the following commuting diagram
\begin{equation}\label{e:fibration}
\xymatrix{
& {\tilde X} \ar[dl]_{\pi'} \ar[dr]^{\pi} \\
B \ar[dr]^{\b} && B' \ar[dl]_{\b'} \\
& \IP^1
}
\end{equation}

The two rational elliptic surfaces $B$ and $B'$ are chosen such that they lie
 in the four-parameter family of rational elliptic surfaces
 described in \cite{Donagi:2000si,Donagi:2004ub}. Both of them
 admit a $\IZ_2$ involution $\t_B$ and $\t_{B'}$ respectively, which lift to a free $\IZ_2$ involution $\t := \t_B \times_{\IP^1} \t_{B'}$ on $\tX$.

\subsection{The Bundle}

To get an $SU(5)$ bundle $V$ on $X$, we construct an $SU(5)$ bundle $\tV$ on $\tX$ together with an action of the involution $\t$ on $\tV$.

Instead of working directly with the bundle $\tV$, in the following we will consider its dual $\tV^*$, since in that case we can apply directly the results of \cite{Donagi:2000sm,Donagi:2004ub}. The bundle $\tV^*$ is constructed as an extension
\begin{equation}\label{e:bdef}
0 \to V_2 \to \tV^* \to V_3 \to 0,
\end{equation}
where $V_2$ and $V_3$ are rank $2$ and $3$ bundles respectively.

The bundles $V_i$ are given by
\begin{equation}
V_i = \pi'^* W_i \otimes \pi^* L_i,
\end{equation}
where the $L_i$ are some line bundles on $B'$ and the $W_i$ are rank $i$ bundles on $B$ given by the Fourier-Mukai transforms $W_i = FM_B (C_i,N_i)$: as usual in the spectral cover construction, the $C_i \subset B$ are curves in $B$ and the $N_i \in Pic(C_i)$ are line bundles over $C_i$.

We choose the following data:
\begin{align}
{\bar C_2} &\in | \co_B ( 2 e_9 +2f)|,\notag \\
C_3 &\in | \co_B ( 3 e_9 + 3 f)|,\notag\\
C_2 &= {\bar C_2} + f_{\infty},\notag\\
N_2 &\in Pic^{3,1} (C_2),\notag\\
N_3 &\in Pic^7 (C_3),\notag\\
L_2 &= \co_B' ( 3 r'),\notag\\
L_3 &= \co_B' ( -2 r').
\end{align}
$f_{\infty}$ is the smooth fiber of $\b$ at $\infty$ containing the four fixed points of $\t_B$, and $Pic^{3,1} (C_2)$ denotes line bundles of degree $3$ over $\bar C_2$ and degree $1$ on $f_{\infty}$. Finally, $r'$ is given by\footnote{For an explanation of the notation see \cite{Bouchard:2005ag}.}
\begin{equation}
r' = e_1' + e_4' -e_5' + e_9' + f'.
\end{equation}

It was shown in \cite{Bouchard:2005ag} that $\tV$ is slope-stable and
invariant under the $\IZ_2$ involution.
Its cohomology was computed in \cite{Bouchard:2005ag}, and it leads to exactly
the MSSM massless particle spectrum, with no exotic particles. Furthermore, it
satisfies the anomaly cancellation condition: $c_2 (T \tX) - c_2 (\tV)$ is an
effective class around which $M5$-branes wrap to cancel the anomaly. This
means that we are in the strongly coupled regime of the heterotic string. It
may also be possible to add a gauge instanton $U$ of small rank in the hidden
sector such that $c_2 (U) = 2 f \times \pt + 6 \pt \times f'$, which would
give a weak coupling vacuum of our model.

In conclusion, the manifold $\tX$ with the $\IZ_2$-invariant stable $SU(5)$ bundle $\tV$ is a good candidate for a realistic compactification of the heterotic string, at least at the level of the massless particle spectrum. The aim of this paper is to compute the tri-linear couplings in the low-energy superpotential of this model.

\section{Massless Spectrum and Couplings of the Effective Theory}\label{s:lep}

In the next subsection we briefly summarize salient features of the MSSM and
compare them with those of the effective theory  for the heterotic
supersymmetric Standard Model. We further  discuss the massless spectrum and
classical superpotential couplings for the specific heterotic string model and
confront it with the features of the MSSM in the subsequent subsections.

\subsection{Minimal Supersymmetric Standard Model}

\begin{table}[tb]
\begin{center}
\begin{tabular}{ccc}\hline\\[-0.5cm]
Superfield & Symbol & Representation\\
\hline
Quarks & $Q$ & $(3,2)_{1/3}$\\
Anti-up & $u$ & $({\bar 3},1)_{-4/3}$ \\
Anti-down & $d$ & $({\bar 3},1)_{2/3}$ \\
Leptons & $L$ & $(1,2)_{-1}$\\
Anti-leptons & $e$ & $(1,1)_2$ \\
Up Higgs & $H$ & $(1,2)_1$ \\
Down Higgs & ${\bar H}$ & $(1,2)_{-1}$ \\
Color Gauge Fields & $G$ & $(8,1)_0$ \\
Weak Gauge Fields & $W^{\pm}, Z, \g$ & $(1,3+1)_0$\\
\hline
\end{tabular}
\end{center}
\caption{The particle spectrum of the MSSM. Only the left-chiral fields are shown. The right-chiral fields have conjugate representations under the gauge group.}
\label{t:MSSM}
\end{table}

The MSSM is the `minimal' $N=1$ supersymmetric extension of the Standard Model
(SM). It has gauge group $SU(3)_C \times SU(2)_L \times U(1)_Y$. The
massless spectrum of $N=1$ superfields and their representations of the SM
gauge group are given in table \ref{t:MSSM}. In particular, the matter chiral
superfields consist of three families of quarks and leptons and one Higgs
doublet pair.

The four-dimensional N=1 supersymmetric Lagrangian  is fully specified by a
K\"ahler potential $K$, a superpotential $W$ and  gauge kinetic functions
$f_i$. The K\"ahler potential is a real function of chiral superfields, while
the superpotential and the gauge kinetic functions are in general holomorphic
functions of chiral superfields. The K\"ahler potential,  among others,
determines the kinetic energy terms for  chiral superfields, the
superpotential carries information on the Yukawa  couplings of these
superfields and the gauge kinetic functions determine  gauge couplings for the
super Yang-Mills sector of the theory.

In the MSSM, the matter particle content is that of  three families of quarks
and leptons and  one Higgs doublet pair, supplemented by the supersymmetric
partners to form chiral superfields.  The K\"ahler potential for these chiral
superfields is chosen so that their kinetic energy is
canonically normalized. On the other hand, in string theory the K\"ahler
potential for matter chiral superfields depends on moduli fields. It also
receives corrections at the higher genus level.  In the heterotic string
context it may be possible to determine these couplings in the classical
limit, by obtaining leading contributions in the  limit of ``large''  K\"ahler
and complex structure  Calabi-Yau moduli. We postpone the study of moduli
dependence of the K\"ahler potential.

In the MSSM  the gauge kinetic functions are  fixed to specific constant
values that  match the experimental values for gauge couplings. On the other
hand, in  string theory gauge kinetic functions are holomorphic functions of
the dilaton (string coupling modulus)  and moduli fields.  In  the heterotic
string theory, at the tree level, the gauge functions are universal and
proportional to the dilaton field.

In the MSSM the superpotential of the matter chiral superfields can have the
following tri-linear (renormalizable) couplings:
\begin{equation}
W=W_1 + W_2,
\end{equation}
where
\begin{align}\label{w1}
W_1 =& \l_l^{ij} e_i L_j \bH + \l_d^{ij} Q_i d_j \bH + \l_u^{ij} Q_i u_j H + \m H \bH,\notag\\
W_2 =& \a_1^{ijk} L_i L_j e_k + \a_2^{ijk} L_i Q_j d_k + \a_3^{ijk} u_i d_j d_k,
\end{align}
$i,j,k$ being generation indices. The terms in $W_1$ conserve baryon and
lepton numbers, while those in $W_2$ do not. The latter couplings can be set
to zero  by imposing  {\em R-parity} symmetry.
In this case, the lepton and baryon violating terms are absent and the lightest
superparticle (LSP) of the MSSM is stable, and is a weakly interacting massive
particle (WIMP); thus a good dark matter candidate. The three first terms of $W_1$
determine  Yukawa couplings between  quarks and Higgs fields, and  leptons and
Higgs fields. After electroweak symmetry breaking, i.e. when Higgs doublets
 $H$ and $\bH$ acquire non-zero vacuum expectation values(VEV's), these terms
give masses to  quarks and leptons. One of the main goal of this paper is to determine the
superpotential Yukawa couplings for quarks and leptons.

The last term of $W_1$ is sometimes referred to as a {\em bare Higgs
$\m$-term}. Since in string theory all the couplings are ``field dependent'',
such bare $\mu$-mass term is absent.

In the original formulation of the  MSSM there are no right-handed neutrinos,
and  the left-handed neutrinos are massless. However, there is now
considerable experimental evidence for neutrino oscillations, which require
non-zero neutrino masses. There are various ways to modify the MSSM in order
to give a mass to the neutrinos. Perhaps the simplest way is to add a Majorana
mass term for the left-handed neutrinos.  In string theory this is typically
hard to achieve --- for a recent study of these issues within heterotic string
orbifolds see \cite{GKLN}.

Another possibility is to introduce right-handed neutrinos $\nu_{Ri}$,
associated with the chiral superfields that are  gauge singlets, i.e.
transform in the $(1,1)_0$ representation of the SM gauge group. These fields
can couple via tri-linear couplings to lepton doublets $L_j$ and the $\bH$
Higgs field:
\begin{equation}
W_{\n} = \lambda^{ij}_\nu \n_{Ri} L_j \bH, \label{wnu}\end{equation} which,
after electroweak symmetry breaking generate  non-zero neutrino masses.

 In fact, we can extend further the MSSM by adding more superfields
$\phi_i$ which are singlets of the SM gauge group,  with  additional
superpotential couplings:
\begin{equation}
W_{\phi} = \lambda^{i}_\phi \ \phi_i H \bH + \b^{ijk} \phi_i \phi_j \phi_k.
\label{wphi}
\end{equation}

If the $\phi_i$ acquire non-zero VEV's the first term induces an effective
$\mu$ parameter for the Higgs doublet pairs: such extensions of the MSSM are
often referred to as next to minimal supersymmetric Standard Model (NMSSM).

In string theory the role of  chiral superfields  which are singlets of the SM
gauge group, such as the fields $\nu_{Ri}$ and $\phi_i$, can potentially be
played by moduli fields. In particular, in our specific heterotic string
theory  construction  we shall  show that the tri-linear couplings of $W_\nu$ and
the first coupling in  $W_\phi$ do exist, and  the fields $\nu_{Ri}$ and
$\phi_j$ are identified with  specific vector bundle moduli. On the other
hand, the second term in $W_\phi$, i.e. the tri-linear couplings of vector
bundle moduli fields $\phi_i$ should be  {\it zero}. In perturbation theory
the moduli do not have self-interactions. It is non-perturbative effects,
such a gauge instantons in the  $E_8$ gauge sector, that are expected to
introduce non-perturbative superpotential for moduli fields, thus allowing
their stabilization. But this is a topic beyond the scope of the paper.

At  energies well below the string scale, supersymmetry is broken.
 At low energies  supersymmetry breaking effects manifest themselves as soft
 supersymmetry breaking mass terms for the SM  matter fields.
 In string theory, supersymmetry breaking and moduli
stabilization can in principle be addressed by studying the strong gauge
dynamics associated with  hidden sector gauge instantons or $M5$ branes
wrapping the effective curves. We do not address this difficult task in this
paper; we shall only focus on the string construction, which at the string
energy scale represents the (stable) four-dimensional supersymmetric solution
of string theory.

\subsection{The MSSM from Heterotic String Theory}

We now review the calculation of the massless spectrum of the heterotic  MSSM
construction.  More details may be found in \cite{Green:1987mn}.

The particle spectrum of the $E_8 \times E_8$ heterotic string consists in the zero-modes of the ten-dimensional Dirac operator, $\ker (\dirac)$. Define
\begin{equation}
Spec = \bigoplus_{q=1,3} H^q(X, \ad V);
\end{equation}
then $\ker (\dirac)$ is given by adding the duals to $Spec$ \cite{Donagi:2004ub}. In fact, $Spec$ gives the left-chiral superfields, while its dual gives the right-chiral superfields.

We compactify the $E_8 \times E_8$ heterotic string theory on a Calabi-Yau
threefold $X$ with fundamental group $\IZ_2$. We also construct an $SU(5)$
bundle $V$ on $X$, which breaks the visible $E_8$ gauge group to an $SU(5)$
gauge group. Then, using a $\IZ_2$ Wilson line, we break the $SU(5)$ grand
unified gauge group down to the MSSM gauge group $SU(3)_C \times SU(2)_L
\times U(1)_Y$.

\begin{table}[tb]
\begin{center}
\begin{tabular}{lcc}\hline
Multiplicity & Representation & Superfield\\
\hline
$1 = h^3(\tX, \co_{\tX})_+$ & $(8,1)_0 \oplus (1,3)_0 \oplus (1,1)_0$ & $G,W^{\pm}, Z, \g$\\
$3 = h^1(\tX, \tV^*)_+$ & $({\bar 3},1)_{-4/3} \oplus (1,1)_2$ & $u,e$\\
$3 = h^1(\tX, \tV^*)_-$ & $(3,2)_{1/3}$ & $Q$\\
$0 = h^1(\tX, \tV)_+$ & $(3,1)_{4/3} \oplus (1,1)_{-2}$ & exotic\\
$0 = h^1(\tX, \tV)_-$ & $({\bar 3},2)_{-1/3}$ & exotic\\
$3 = h^1(\tX, \wedge^2 \tV^*)_+$ & $({\bar 3},1)_{2/3}$ & $d$\\
$3+n = h^1(\tX, \wedge^2 \tV^*)_-$ & $(1,{\bar 2})_{-1}$ & $L,\bH$\\
$0 = h^1(\tX, \wedge^2 \tV)_+$ & $({3},1)_{-2/3}$ & exotic\\
$n = h^1(\tX, \wedge^2 \tV)_-$ & $(1,2)_{1}$ & $H$\\
$51 = h^1 (\tX, \ad \tV)_+$ & $(1,1)_0$ & $\phi, \n$\\
\hline
\end{tabular}
\end{center}
\caption{The particle spectrum of the low-energy $SU(3)_C \times SU(2)_L \times U(1)_Y$ theory. Notice that all exotic particles come with $0$ multiplicity, and that the spectrum include $n$ copies of Higgs conjugate pairs, where $n=0,1,2$.}
\label{t:lespec}
\end{table}

The resulting low-energy superfields are given by the decomposition of $Spec$ under
the above symmetry breaking pattern. In particular, the multiplicity of the representations
 of the low-energy MSSM gauge group are given by the dimensions of the invariant and anti-invariant
  parts of some cohomology groups. For the case under consideration, the decomposition of $Spec$ and
   the associated cohomology groups has been worked out in \cite{Donagi:2004ub}. The resulting
   low-energy spectrum is shown in table \ref{t:lespec}.

We computed the required cohomology groups in \cite{Bouchard:2005ag}, and
found the multiplicities also presented in table \ref{t:lespec}. Notice that
the low-energy spectrum has three generations of quarks and leptons, no exotic
particles, $0,1$ or $2$ pairs of Higgs, and $51$ vector bundle moduli
fields.\footnote{In fact we did not compute the dimension of $H^1(\tX, \ad
\tV)$ in \cite{Bouchard:2005ag}; we simply gave an estimate of its dimension
from simple parameter counting. We will study this cohomology group in more
detail in section \ref{s:moduli}.}

\subsection{Superpotential}

The main focus of this paper is to compute the superpotential $W$ of the
model. More precisely, we want to determine which terms  in  $W_1$, $W_2$,
$W_{\n}$ and $W_{\phi}$ are non-vanishing. Computing the exact numerical
coefficients and their explicit dependence on the (vacuum expectation) values
of the moduli is harder and we shall not do that explicitly. In addition our
calculation is done only in the classical limit, and thus the quantum
(world-sheet instanton) effects are not included.

We have seen that massless superfields correspond to equivalence classes in
some cohomology groups of some bundles over our Calabi-Yau threefold $X$. In
other words, we can associate to each superfield a $\bar{\partial}$-closed
$(0,1)$-form $\Phi_i$ taking values in some bundle over $X$. Compactifying
heterotic string theory on $X$ yields cubic terms in the superpotential of the
four-dimensional effective action. The coefficients of these terms are given
by the unique way of extracting a complex number out of the three associated
$(0,1)$-forms $\Phi_i$, that is, by wedging the three  $(0,1)$-forms and the
holomorphic volume $(3,0)$-form $\Omega$ of $X$ and integrating over $X$:
 \begin{equation} \lambda_ {ijk}\sim \int_X\ \Omega\wedge
\Phi_i\wedge\Phi_j \wedge \Phi_k\, .
 \end{equation}
 Note that this is only a term determined at  the tree level of sigma model perturbation,
that is in the ``large volume limit''; these
coefficients can receive corrections due to worldsheet instantons
 but we will not compute them in this paper.

In cohomological language, the coefficients are given by the
 images in $\IC$ of some triple pairings of cohomology groups.
  Thus, to compute the coefficients in the superpotential we must first find all
   possible triple pairings of cohomology
   groups\footnote{We do not consider the group $H^3 (\tX, \co_{\tX})$,
   since it corresponds to the gauge connections and thus determine non-Abelian gauge
   couplings; our focus is on the superpotential tri-linear couplings of the  matter chiral
   superfields.}
    in table \ref{t:lespec} mapping into $\IC$; these are the cubic terms that may appear in the superpotential of the four-dimensional effective action. Then we must show whether these pairings vanish or not.

Using the multiplicities given in table \ref{t:lespec}, in particular the fact that all exotic particles have multiplicity zero, we find the following allowed triple pairings\footnote{In the following we will sometimes suppress the $\tX$ in the cohomology groups for clarity.}:
\begin{align}\label{e:tricoup}
{\rm \bf (d)}& & H^1 (\wedge^2 \tV^*)^{(3,3+n)} \times H^1 (\wedge^2 \tV^*)^{(3,3+n)} \times H^1 (\tV^*)^{(3,3)} &\to H^3 (\wedge^5 \tV^*)^{(1,0)} \simeq \IC,\notag\\
{\rm \bf (u)}& &H^1 ( \tV^*)_+^{(3,0)} \times H^1 ( \tV^*)_-^{(0,3)} \times H^1 (\wedge^2 \tV)_-^{(0,n)} &\to H^3 (\co)^{(1,0)} \simeq \IC,\notag\\
{\rm \bf (\mu)}& &H^1 ( \ad \tV)_+^{(51,0)} \times H^1 ( \wedge^2 \tV^*)_-^{(0,3+n)} \times H^1 (\wedge^2 \tV)_-^{(0,n)} &\to H^3 (\co)^{(1,0)} \simeq \IC,\notag\\
{\rm \bf (\phi)}& &H^1 ( \ad \tV)_+^{(51,0)} \times H^1 ( \ad \tV)_+^{(51,0)} \times H^1 (\ad \tV)_+^{(51,0)} &\to H^3 (\co)^{(1,0)} \simeq \IC,\notag\\
\end{align}
where the pairings are given by cup product and wedge product, and $n=0,1,2$ (note that for $n=0$
 the {\bf (u)} and {\bf ($\m$)} pairings vanish identically since there is no Higgs pair).
 The superscripts $(x,y)$ mean that the invariant part of the cohomology
  group has dimension $x$, while the anti-invariant part has dimension $y$.
   Each of these pairings correspond to various cubic couplings in the superpotential;
    they can be read off from the associated superfields presented in table \ref{t:lespec}.
     The names we gave to the triple pairings are of physical significance,
     as will be explained in more detail in section \ref{s:pheno}.
      Jumping ahead a little, let us simply say that the {\bf (d)} pairing
corresponds to couplings of down sector quarks  and charged lepton sector to
 the down-Higgs doublet, and also to the potential R-parity violating couplings of $W_2$ in \eqref{w1}.  The {\bf (u)}
       pairing is related to the Yukawa couplings of the up-sector quarks to the up-Higgs doublet.
        The {\bf ($\mu$)}
        pairing corresponds to the moduli-dependent Higgs $\m$-terms
(the first term of $W_\phi$  in \eqref{wphi})        and potential neutrino
mass terms ($W_\nu$ in  \eqref{wnu}). Finally, the {\bf ($\phi$)} pairing corresponds to the tri-linear couplings of the vector bundle moduli
         (the second term of $W_\phi$ in \eqref{wphi}).

\section{Vector Bundle Moduli}\label{s:moduli}

In this section we study the vector bundle moduli space $H^1 (\ad \tV)$,
which enters into the calculation of the tri-linear couplings.

In fact, we only need its
invariant subspace, which we denote by $H^1 (\ad \tV)_+$, as shown in table
\ref{t:lespec}. Note that $\ad \tV$ is defined to be the traceless part of
$\tV \otimes \tV^*$. But since $H^1 (\co) = 0$, we can identify
\begin{equation}
H^1 (\ad \tV)_+ \simeq H^1 (\tV \otimes \tV^*)_+ .
\end{equation}

Before doing anything, we must understand the cohomology $H^* (V_2 \otimes V_3^*)_+$. We computed in \cite{Bouchard:2005ag} that\footnote{Note that in \cite{Bouchard:2005ag} we stated that the invariant subspace of the 90-dimensional cohomology space is 50-dimensional, but a careful analysis shows that it is rather 45-dimensional.}
\begin{equation}
h^1 (V_2 \otimes V_3^*)_+ = 45.
\end{equation}
Using similar techniques, it is straightforward to show that
\begin{equation}
h^0(V_2 \otimes V_3^*)_+ = h^2(V_2 \otimes V_3^*)_+ =h^3(V_2 \otimes V_3^*)_+ = 0.
\end{equation}

We are now ready to attack the computation of $H^1 (\tV \otimes \tV^*)_+$. Recall that
\begin{equation}\label{e:esdef}
0 \to V_2 \to \tV^* \to V_3 \to 0
\end{equation}
implies the long exact sequence in cohomology
\begin{equation}\label{e:estbt}
\ldots \to H^0 (V_3) \to H^1 (V_2) \to H^1 (\tV^*) \to H^1 (V_3) \to H^2 (V_2) \to \ldots
\end{equation}
Tensoring \eqref{e:esdef} with $V_3^*$, taking the long exact sequence in cohomology (keeping only the invariant subspaces of the cohomology groups) and using the results above we obtain the exact sequence
\begin{multline}
0 \to H^0 (\tV^* \otimes V_3^*)_+ \to H^0 (V_3 \otimes V_3^*)_+ \xrightarrow{d_1} H^1 (V_2 \otimes V_3^*)_+ \\
\to H^1 (\tV^* \otimes V_3^*)_+ \to H^1 (V_3 \otimes V_3^*)_+ \to 0.
\end{multline}
Only the trace part of $V_3 \otimes V_3^*$ contributes to $H^0$, and so $h^0 (V_3 \otimes V_3^*)_+ = h^0 (\co)_+ = 1$. Thus, the map $d_1$ that we identified in the long exact sequence is simply multiplication of constant sections by the invariant extension class of our bundle in $H^1 (V_2 \otimes V_3^*)_+$. Since our bundle is a non-trivial extension, the map $d_1$ is non-zero, and therefore must have rank $1$. Hence
\begin{equation}
H^0 (\tV^* \otimes V_3^*)_+ =0,
\end{equation}
and the sequence reduces to
\begin{equation}\label{e:seqred}
0 \to H^0 (V_3 \otimes V_3^*)_+ \rightarrow H^1 (V_2 \otimes V_3^*)_+
\to H^1 (\tV^* \otimes V_3^*)_+ \to H^1 (V_3 \otimes V_3^*)_+ \to 0.
\end{equation}

From \eqref{e:seqred} we see that $H^1 (\tV^* \otimes V_3^*)_+$ has a filtration $F^0 \supseteq F^1 \supseteq \{0\}$, with $F^0 = H^1 (\tV^* \otimes V_3^*)_+$ and
\begin{equation}
F^1 = \frac{H^1 (V_2 \otimes V_3^*)_+}{H^0 (V_3 \otimes V_3^*)_+}.
\end{equation}
Its associated graded vector space is
\begin{equation}
\frac{H^1 (V_2 \otimes V_3^*)_+}{H^0 (V_3 \otimes V_3^*)_+} \oplus \frac{H^1 (\tV^* \otimes V_3^*)_+}{H^1 (V_2 \otimes V_3^*)_+/H^0 (V_3 \otimes V_3^*)_+} \simeq
\frac{H^1 (V_2 \otimes V_3^*)_+}{H^0 (V_3 \otimes V_3^*)_+} \oplus H^1 (V_3 \otimes V_3^*)_+.
\end{equation}

Let us now tensor the exact sequence \eqref{e:esdef} with $V_2^*$ and take the long exact sequence in cohomology (keeping again only the invariant subspaces). Using Serre duality, we obtain
\begin{equation}
0 \to H^i (V_2 \otimes V_2^*)_+ \to H^i (\tV^* \otimes V_2^*)_+ \to 0,
\end{equation}
for $i=0,1$, which implies that
\begin{equation}
H^i (\tV^* \otimes V_2^*)_+ \simeq H^i (V_2 \otimes V_2^*)_+,
\end{equation}
for $i=0,1$.

Consider now the short exact sequence dual to \eqref{e:esdef}, and tensor it with $\tV^*$. Taking the associated long exact sequence in cohomology we obtain
\begin{multline}
0 \to H^0 (\tV \otimes \tV^*)_+ \to H^0 (\tV^* \otimes V_2^*)_+ \xrightarrow{d_2} H^1 (\tV^* \otimes V_3^*)_+ \\
\to H^1 (\tV \otimes \tV^*)_+ \to H^1 (\tV^* \otimes V_2^*)_+ \xrightarrow{\d} H^2 (\tV^* \otimes V_3^*)_+ \to \ldots
\end{multline}
We know that $h^0 (\tV^* \otimes V_2^*)_+ = h^0 (V_2 \otimes V_2^*)_+ = 1$. Thus $h^0 (\tV \otimes \tV^*)_+$ is $0$ or $1$. But it cannot be $0$, since $\tV \otimes \tV^* = \co \oplus (\tV \otimes \tV^*)_{\rm traceless}$ and $h^0(\co)_+=1$. Thus $h^0 (\tV \otimes \tV^*)_+=1$, which implies that ${\rm rank}(d_2) = 0$ and the sequence splits.

Now we must understand the coboundary map $\d$. It is given by cup product with the invariant extension class
\begin{equation}
H^1 (\tV^* \otimes V_2^*)_+ \times H^1 (V_2 \otimes V_3^*)_+ \to H^2 (\tV^* \otimes V_3^*)_+.
\end{equation}
But since $H^1 (\tV^* \otimes V_2^*)_+ \simeq H^1 (V_2 \otimes V_2^*)_+$, the image must lie in $H^2 (V_2 \otimes V_3^*)_+$, which is zero. Thus $\delta = 0$, and we obtain the short exact sequence
\begin{equation}
0 \to H^1 (\tV^* \otimes V_3^*)_+ \to H^1 (\tV \otimes \tV^*)_+ \to H^1 (\tV^* \otimes V_2^*)_+ \to 0 .
\end{equation}

Using all these results, we conclude that the space of vector bundle moduli $H^1 (\tV \otimes \tV^*)_+$ has a filtration $G^0 \supseteq G^1 \supseteq G^2 \supseteq \{0\}$, with $G^0 = H^1 (\tV \otimes \tV^*)_+$, $G^1 = H^1 (\tV^* \otimes V_3^*)_+$ and
\begin{equation}
G^2 = \frac{H^1 (V_2 \otimes V_3^*)_+}{H^0 (V_3 \otimes V_3^*)_+}.
\end{equation}
Its associated graded vector space is given by
\begin{equation}\label{e:gvs}
\frac{H^1 (V_2 \otimes V_3^*)_+}{H^0 (V_3 \otimes V_3^*)_+} \oplus H^1 (V_3 \otimes V_3^*)_+\oplus H^1 (V_2 \otimes V_2^*)_+,
\end{equation}
where we used the fact that
\begin{equation}
H^1 (V_2 \otimes V_2^*)_+ \simeq \frac{H^1 (\tV \otimes \tV^*)_+}{H^1 (\tV^* \otimes V_3^*)_+}.
\end{equation}

This result is easy to understand. The first factor of \eqref{e:gvs} is the invariant extension space, quotiented by a one-dimensional space. This simply means that we are free to choose any invariant extension class for our bundle $\tV$, but rescaling does not change the bundle. The second factor corresponds to moduli coming from the vector bundle $V_3$, while the third factor corresponds to moduli coming from the vector bundle $V_2$.

\subsection{Dimension}

In order to get the dimension of the space of vector bundle moduli $H^1 (\tV \otimes \tV^*)_+$ we simply have to add up the dimensions of the three spaces in the associated graded vector space \eqref{e:gvs}. First, we computed in \cite{Bouchard:2005ag} that the invariant subspace of $H^1 (V_2 \otimes V_3^*)$ is 45-dimensional. Thus the first factor of \eqref{e:gvs} is 44-dimensional. We now compute the dimension of the two other cohomology spaces in \eqref{e:gvs}. Let us first study $H^1 (V_3 \otimes V_3^*)_+$.

To start with, we can use a Leray spectral sequence to show that
\begin{equation}
H^1 (\tX, V_3 \otimes V_3^*) \simeq H^1 (B, W_3 \otimes W_3^*).
\end{equation}
Hence we must compute the cohomology $H^* (B, W_3 \otimes W_3^*)$ on the rational elliptic surface $B$. Only the trace part contributes to $H^0$, and we obtain
\begin{equation}
h^0 (B, W_3 \otimes W_3^*) = 1.
\end{equation}
Using Serre duality and a Leray spectral sequence, we also find
\begin{align}
H^2 (B, W_3 \otimes W_3^*) &\simeq H^0 (B, W_3 \otimes W_3^* \otimes \co(-f) )^*\notag\\
 &\simeq H^0 (\IP^1, \co(-1) \otimes \b_* (W_3 \otimes W_3^*) )^* = 0.
\end{align}
To compute the remaining cohomology group $H^1(B, W_3 \otimes W_3^*)$, we use the Hirzebruch-Riemann-Roch theorem (or index theorem). For a rational elliptic surface $B$, we have that
\begin{equation}
{\rm td}(B)= 1 + f + \pt.
\end{equation}
The Chern character of $W_3$ was computed in \cite{Bouchard:2005ag}, which gives
\begin{equation}
{\rm ch} (W_3 \otimes W_3^*) = {\rm ch} (W_3) \cdot {\rm ch} (W_3^*) = (3+f-3 \pt) \cdot (3-f-3 \pt) = 9 - 18 \pt .
\end{equation}
Then, Hirzebruch-Riemann-Roch tells us that
\begin{equation}
h^0 - h^1 + h^2 = 1 - h^1 = \deg [ (1+f+\pt) \cdot (9 - 18 \pt) ]_2 = -9,
\end{equation}
which implies that
\begin{equation}
h^1 (\tX, V_3 \otimes V_3^*) = h^1 (B, W_3 \otimes W_3^*) = 10.
\end{equation}

We can do exactly the same computation for $V_2 \otimes V_2^*$, and we obtain
\begin{equation}
h^1 (\tX, V_2 \otimes V_2^*) = h^1 (B, W_2 \otimes W_2^*) = 9.
\end{equation}

Now we must find the dimension of the invariant subspaces. In order to do so, let us give a geometrical description of these cohomology groups. $H^1 (B, W_3 \otimes W_3^*)$ (and similarly for $W_2$) is the space of vector bundle moduli for $W_3$. From the spectral cover construction, we see that it splits into two components; the dimension of the (projectivization) of the linear system of which the spectral curve $C_3$ is an element, and the genus of the curve $C_3$. More precisely, in this case the linear system is \cite{Donagi:2000sm}
\begin{equation}
|3e + 3f| = H^0 (B, \co(3e+3f)) \simeq H^0 (\IP^1, \co(3)) \oplus H^0 (\IP^1, \co(1)) \oplus H^0 (\IP^1, \co), 
\end{equation}
which has dimension $7$. The arithmetic genus of $C_3$ is $4$, and thus
\begin{equation}
h^1 (B, W_3 \otimes W_3^*) = 4 + 7 -1 = 10,
\end{equation}
which is indeed correct. In the case of $W_2$, the linear system is given by
\begin{equation}
|2e + 3f| = H^0 (B, \co(2e+3f)) \simeq H^0 (\IP^1, \co(3)) \oplus H^0 (\IP^1, \co(1)),
\end{equation}
which has dimension $6$, and the curve $C_2$ has arithmetic genus $4$. Hence
\begin{equation}
h^1 (B, W_2 \otimes W_2^*) = 4 + 6 -1 = 9.
\end{equation}

Using the decomposition of the linear systems in terms of cohomology groups on $\IP^1$, we can find the dimension of the invariant and anti-invariant parts. The $7$-dimensional space $H^0 (B, \co(3e+3f))$ breaks into a $4$-dimensional invariant subspace and a $3$-dimensional anti-invariant subspace, while the $6$-dimensional space $H^0 (B, \co(2e+3f))$ breaks into a $3$-dimensional invariant subspace and a $3$-dimensional anti-invariant suspace. The remaining step is to find the genus of the quotient curves.

As described in \cite{Donagi:2000sm}, the involution $\a_B$ on $B$ fixes the fiber $f_0$ above $0 \in \IP^1$ pointwise, and has $4$ isolated fixed points in the fiber $f_{\infty}$ above $\infty \in \IP^1$. An invariant curve $C_3 \in |3e+3f|_+$ intersects $f_{\infty}$ at three of these four fixed points, and intersects $f_0$ in three points. Thus the action of $\a_B$ on $C_3$ has $6$ fixed points. Using Hurwitz' theorem, we find
\begin{equation}
2 g (C_3 / \a_B) - 2 = \frac{1}{2} \left ( 2 g (C_3) - 2 - 6 \right) = \frac{1}{2} \left ( 8 -2 -6 \right) = 0,
\end{equation}
which implies
\begin{equation}
g (C_3 / \a_B) = 1 .
\end{equation}

In the case of $C_2 \in |2e + 3f|_+$, a generic curve consists in a reducible curve which contains $f_{\infty}$ plus a curve ${\bar C_2} \in |2e + 2f|_+$. The quotient of such a curve is an elliptic curve plus a rational curve attached to it at one point, which has arithmetic genus $1$. In other words, the action of $\a_B$ on $C_2$ has $4$ fixed points in $f_{\infty}$, and two fixed points at the intersection points of $C_2$ and $f_0$. Thus by Hurwitz' theorem
\begin{equation}
 2 g (C_2 / \a_B) - 2 = \frac{1}{2} \left ( 2 g (C_2) - 2 - 6 \right) = \frac{1}{2} \left ( 8 -2 -6 \right) = 0,
\end{equation}
which again implies
\begin{equation}
g (C_2 / \a_B) = 1 .
\end{equation}

Putting all these results together, we find that the invariant parts of the cohomology groups have dimensions
\begin{align}
h^1 (V_3 \otimes V_3^*)_+ &= 1 + 4 - 1 = 4,\notag\\
h^1 (V_2 \otimes V_2^*)_+ &= 1 + 3 - 1 = 3.
\end{align}

Therefore, we conclude that the space of vector bundle moduli has dimension
\begin{equation}
h^1 (V \otimes V^*)_+ = 45-1+4+3 = 51.
\end{equation}

\section{Computation of the Triple Pairings}\label{s:comp}

We now compute the tri-linear couplings given by  \eqref{e:tricoup}.

\subsection{{\bf (d)} Triple Pairing}

Let us start by analyzing the {\bf (d)} triple pairing, given by
\begin{equation}\label{e:coupR}
{\rm \bf (d)}~~~~ H^1 (\wedge^2 \tV^*)^{(3,3+n)} \times H^1 (\wedge^2
\tV^*)^{(3,3+n)} \times H^1 (\tV^*)^{(3,3)} \to \IC.
\end{equation}
First, recall from \cite{Bouchard:2005ag} that $H^1 (\tV^*) \simeq H^1 (V_2)$, and that we have a long exact sequence
\begin{equation}\label{e:es}
0 \xrightarrow{} H^1 (\wedge^2 V_2) \xrightarrow{} H^1 (\wedge^2 \tV^*) \xrightarrow{} H^1 (V_2 \otimes V_3) \xrightarrow{M^T} H^2 (\wedge^2 V_2) \xrightarrow{} H^2 (\wedge^2 \tV^*) \xrightarrow{} 0.
\end{equation}
Thus $H^1 (\wedge^2 \tV^*)$ has a filtration $F^0 \supseteq F^1 \supseteq \{ 0 \}$, with $F^0 = H^1 (\wedge^2 \tV^*)$ and $F^1 = H^1 (\wedge^2 V_2)$. Its associated graded vector space is
\begin{equation}\label{e:gvs2}
H^1 (\wedge^2 V_2) \oplus \frac{H^1 (\wedge^2 \tV^*)}{H^1 (\wedge^2 V_2)} \simeq H^1 (\wedge^2 V_2) \oplus \ker M^T.
\end{equation}

If we restrict both $H^1 (\wedge^2 \tV^*)$ in \eqref{e:coupR} to their subspaces $H^1 (\wedge^2 V_2)$, the {\bf (d)} pairing becomes
\begin{equation}
H^1 (\wedge^2 V_2) \times H^1 (\wedge^2 V_2) \times H^1 (V_2) \to H^3 (\wedge^5 V_2) = 0,
\end{equation}
which vanishes since $V_2$ has rank $2$ and thus $\wedge^i V_2 = 0$ for $i > 2$.

Suppose now that one $H^1 (\wedge^2 \tV^*)$ factor lives in its $H^1 (\wedge^2 V_2)$ subspace and that the other one lives in the quotient space
\begin{equation}
\frac{H^1 (\wedge^2 \tV^*)}{H^1 (\wedge^2 V_2)} \simeq \ker M^T \subseteq H^1 (V_2 \otimes V_3).
\end{equation}
Then, the {\bf (d)} pairing becomes
\begin{equation}
H^1 (\wedge^2 V_2) \times H^1 (V_2 \otimes V_3) \times H^1 (V_2) \to H^3 (\wedge^4 V_2 \otimes V_3) = 0.
\end{equation}

Thus both factors must be in their quotient spaces. But even then, we find
\begin{equation}
H^1 (V_2 \otimes V_3) \times H^1 (V_2 \otimes V_3) \times H^1 (V_2) \to H^3 (\wedge^3 V_2 \otimes \wedge^2 V_3) = 0,
\end{equation}
from which we conclude that the {\bf (d)} pairing vanishes identically for $n=0,1,2$.

\subsection{{\bf (u)} Triple Pairing}

We now turn to the pairing
\begin{equation}
{\rm \bf (u)}~~~~H^1 ( \tV^*)_+^{(3,0)} \times H^1 ( \tV^*)_-^{(0,3)} \times H^1 (\wedge^2 \tV)_-^{(0,n)} \to \IC,
\end{equation}
or equivalently
\begin{equation}\label{e:ypair}
H^1 ( \tV^*)_+^{(3,0)} \times H^1 ( \tV^*)_-^{(0,3)} \to H^2 (\wedge^2 \tV^*)_-^{(0,n)}.
\end{equation}

Recall that $H^1 (\tV^*)_{\pm} \simeq H^1 (V_2)_{\pm}$, and that from the long exact sequence \eqref{e:es} the map
\begin{equation}
\a: H^2 (\wedge^2 V_2)_-^{(0,8)} \to H^2 ( \wedge^2 \tV^*)_-^{(0,n)}
\end{equation}
is surjective. In fact, we can identify $H^2 ( \wedge^2 \tV^*)_- \simeq \coker M^T$. Thus let us first analyze the pairing
\begin{equation}\label{e:pairY}
H^1 ( V_2)_+^{(3,0)} \times H^1 ( V_2)_-^{(0,3)} \to H^2 (\wedge^2 V_2)_-^{(0,8)}
\end{equation}
and then use $\a$ to project to $H^2 ( \wedge^2 \tV^*)_-^{(0,n)}$, i.e. mod out by the anti-invariant part of $\Ima M^T$, which we denote by $(\Ima M^T)_-$.

As explained in \cite{Bouchard:2005ag,Donagi:2004ub} there is a natural identification $H^1 ( V_2) \simeq S_x^4 \oplus y S_x^1$ and $H^2 (\wedge^2 V_2) \simeq S_x^6 \oplus y S_x^4 \otimes (S_t^1)^*$. Thus we can write an explicit basis for the vector spaces involved in the pairing \eqref{e:pairY}:\footnote{Recall that \cite{Bouchard:2005ag} the $\IZ_2$ action is given by $t_0 \mapsto t_0$, $t_1 \mapsto - t_1$, $x_0 \mapsto x_0$, $x_1 \mapsto -x_1$, $y \mapsto y$.}
\begin{align}\label{e:basis}
H^1 ( V_2)_+: &\{ x_0^3, x_0 x_1^2; y x_0 \}, \notag\\
H^1 (V_2)_-: &\{ x_0^2 x_1, x_1^3; y x_1 \},\notag\\
H^2 (\wedge^2 V_2)_-: &\{ x_0^5 x_1, x_0^3 x_1^3, x_0^1 x_1^5; y t_0 x_0^3 x_1, y t_0 x_0 x_1^3; y t_1 x_0^4, y t_1 x_0^2 x_1^2, y t_1 x_1^4 \}.
\end{align}
For instance, an element of $H^1 (V_2)_+$ is given by a polynomial
\begin{equation}
a_{3 0} x_0^3 + a_{12} x_0 x_1^2 + a_{10} y x_0,
\end{equation}
where the subscripts denote the powers of $x_0$ and $x_1$ respectively, or simply by a vector
\begin{equation}
(a_{30}, a_{12}, a_{10} ).
\end{equation}

Using this description of the cohomology groups we can write down the map
\eqref{e:pairY} explicitly. It is given by
\begin{multline}\label{e:expmap}
(a_{30}, a_{12}, a_{10} ) \times (a_{21},a_{03},a_{01}) \mapsto (b_{51},b_{33},b_{15},b_{31},b_{13},b_{40},b_{22},b_{04} ) \\
= (a_{30} a_{21}, a_{30} a_{03} + a_{12} a_{21}, a_{12} a_{03}, a_{30} a_{01} + a_{10} a_{21}, a_{10} a_{03} + a_{12} a_{01}, 0, 0, 0).
\end{multline}
The image lives in the five-dimensional subspace of $H^2 (\wedge^2 V_2)_-$ given by evaluating at $[t_0:t_1] = [1:0]$.

Now to get the {\bf (u)} pairing, we must project to $H^2 ( \wedge^2 \tV^*)_-^{(0,n)}$, that is we must quotient by $(\Ima M^T)_-$. In other words, we must determine whether the above five-dimensional subspace (or any subspace thereof) lies in $(\Ima M^T)_-$ --- which would imply that (some of) the {\bf (u)} pairings vanish --- or not.

Recall from \cite{Bouchard:2005ag} that the map
\begin{equation}
M^T: H^1 (V_2 \otimes V_3) \to H^2 (\wedge^2 V_2),
\end{equation}
given by evaluating at the invariant extension class of $\tV$, can be expressed in matrix form. More precisely, the $18 \times 17$ matrix $M$ has the form
\begin{equation}
\left(
\begin{array}{cccccc|cccccc}
&A_-&&0&&&&&\\
0 &&B_-&&&&&&\\
\hline
&&&&&&&~~~&A_-&~~~&\\
&&&&&&&~~~&B_-&~~~&\\
\end{array}
\right),
\end{equation}
where $0$ means a column of zeroes, the top-left block has dimension $9 \times 9$ and the bottom-right block has dimension $9 \times 8$. In order to get $n$ pairs of Higgs (for $n=0,1,2$), we require that the matrix $M$ has rank $17-n$. From simple dimension counting this implies that $\Ima M^T$ has codimension $n$ as required. In fact, since we specifically want the anti-invariant part of $\Ima M^T$ to have codimension $n$, we need a slightly stronger condition on $M$. We demand that the bottom-right $9 \times 8$ block has rank $8-n$, while the top-left $9 \times 9$ block has rank $9$. It was shown in \cite{Bouchard:2005ag} that on the loci in moduli space where the bottom-right block has rank $8-n$ for $n=1,2$, requiring that the top-left block has rank $9$ is an open condition, but not empty. So, on these loci, the solution space is a dense open subset. Thus we can get almost any subspace of codimension $n$ (for $n=1,2$) in $H^2 ( \wedge^2 V_2)_-^{(0,8)}$ as $(\Ima M^T)_-$.

In particular, for a generic choice of invariant extension class satisfying the above condition on the rank of $M$, the five-dimensional subspace of $H^2 ( \wedge^2 V_2)_-^{(0,8)}$ described in \eqref{e:expmap} (or any subspace thereof) will not lie in $(\Ima M^T)_-$. Hence, after quotienting by $(\Ima M^T)_-$ we obtain non-zero values, and the {\bf (u)} pairings \eqref{e:ypair} are non-zero.

To make things more transparent, let us look first at the one Higgs case. Then, we can express the {\bf (u)} pairing \eqref{e:ypair} between the two $3$-dimensional spaces as a $3 \times 3$ matrix. From the explicit description of \eqref{e:expmap}, we see that we obtain a symmetric $3 \times 3$ matrix with the bottom-right entry being zero:
\begin{equation}\label{e:yukmat}
\begin{pmatrix}
a & b & c\\
b & d & e\\
c & e & 0
\end{pmatrix}.
\end{equation}
To obtain this matrix, consider elements of the first $3$-dimensional space in the pairing \eqref{e:pairY} as column vectors, elements of the second $3$-dimensional space as rwo vectors, and multiply them to get a $3 \times 3$ matrix. Then set the bottom-right entry to zero and symmetrize the matrix. This reformulation of \eqref{e:expmap} is an explicit version of the map \eqref{e:pairY}. Here each of the entries $a,b,c,d,e$ is an element of $H^2(\wedge^2 V_2)$, which by \eqref{e:basis} can be written as a linear combination of the 8 basic
monomials. By reading off the coefficients, the $3 \times 3$ matrix of polynomials \eqref{e:yukmat} can therefore be interpreted as a set of eight $3 \times 3$ matrices with
scalar entries. The conclusion of the previous discussion is that our original coupling \eqref{e:ypair} is given by a generic linear combination of these
eight matrices.

In other words, what we obtained above is that generically, the coefficients in the matrix \eqref{e:yukmat} are non-zero, and so the matrix has rank $3$.

For the $n=2$ case, we obtain two $3 \times 3$ matrices of the form \eqref{e:yukmat} (one for each pair of Higgs). Generically, they both have rank $3$.


\subsection{{\bf ($\mu$)} Triple Pairing}

The {\bf ($\mu$)} pairing is given by
\begin{equation}
{\rm \bf (\m)}~~~~ H^1 ( \ad \tV)_+^{(51,0)} \times H^1 ( \wedge^2
\tV^*)_-^{(0,3+n)} \times H^1 (\wedge^2 \tV)_-^{(0,n)} \to \IC,
\end{equation}
which can be rewritten as
\begin{equation}\label{e:pairmod}
H^1 (\tV \otimes \tV^*)_+^{(51,0)} \times H^1 (\wedge^2 \tV^*)_-^{(0,3+n)} \to
H^2 (\wedge^2 \tV^*)_-^{(0,n)}.
\end{equation}

Recall that $H^1 (\wedge^2 \tV^*)_-$ has a filtration $F^0 \supseteq F^1 \supseteq \{ 0 \}$, with $F^0 = H^1 (\wedge^2 \tV^*)_-$, $F^1 = H^1 (\wedge^2 V_2)_-$ and $F^0 / F^1 \simeq \ker M^T$.

Recall also that $H^1 (\tV \otimes \tV^*)_+$ has a filtration $G^0 \supseteq G^1 \supseteq G^2 \supseteq \{0\}$, with $G^0 = H^1 (\tV \otimes \tV^*)_+$, $G^1 = H^1 (\tV^* \otimes V_3^*)_+$,
\begin{equation}
G^2 = \frac{H^1 (V_2 \otimes V_3^*)_+}{H^0 (V_3 \otimes V_3^*)_+},
\end{equation}
$G^1 / G^2 \simeq H^1 (V_3 \otimes V_3^*)_+$ and $G^0 / G^1 \simeq H^1(V_2 \otimes V_2^*)_+$.

Finally, recall that $H^2 (\wedge^2 \tV^*)_- \simeq  \coker M^T$.

Let us first restrict $H^1 (\wedge^2 \tV^*)_-$ to its $F^1 =  H^1 (\wedge^2 V_2)_-$ subspace. Since $(\Ima M^T)_- \subset H^2 (\wedge^2 V_2)_-$ and $H^2 (\wedge^2 \tV^*)_- \simeq  \coker M^T$, we can replace $H^2 (\wedge^2 \tV^*)_-$ in the pairing \eqref{e:pairmod} by $H^2 (\wedge^2 V_2)_-$, keeping in mind that we want to quotient by $(\Ima M^T)_-$ afterwards. After rearrangement, the pairing \eqref{e:pairmod} becomes
\begin{equation}
H^1 (\wedge^2 V_2)_- \times H^1 (\wedge^2 V_2^*)_- \to H^2 (\tV \otimes \tV^*).
\end{equation}
But $\wedge^2 V_2$ and $\wedge^2 V_2^*$ are dual line bundles, and so the pairing must be
\begin{equation}
H^1 (\wedge^2 V_2)_- \times H^1 (\wedge^2 V_2^*)_- \to H^2 (\Tr (\tV \otimes \tV^*)) = H^2(\co) = 0,
\end{equation}
which vanishes.

Hence, only the quotient space $F^0 / F^1 \simeq \ker M^T$ of $F^0 = H^1 (\wedge^2 \tV^*)_-$ may give non-zero pairings. Restricting to the subspace
\begin{equation}
G^2 = \frac{H^1 (V_2 \otimes V_3^*)_+}{H^0 (V_3 \otimes V_3^*)_+} \subseteq H^1 (\tV \otimes \tV^*)_+
\end{equation}
we get the remaining {\bf ($\m$)} pairings
\begin{equation}\label{e:muterms}
{\left (\frac{H^1(V_2 \otimes V_3^*)_+}{H^0 (V_3 \otimes V_3^*)_+}\right)}^{(44,0)} \times (\ker M^T)^{(0,1+n)} \to (\coker M^T)^{(0,n)},
\end{equation}
which are generically non-zero.

We conclude that in the one Higgs case, only $2$ of the vector bundle moduli have non-vanishing {\bf ($\mu$)} couplings, while in the two Higgs case only $6$ of the vector bundle moduli have non-vanishing {\bf ($\mu$)} couplings. For $n=1$ (respectively $n=2$), one coupling (respectively four) involves a down Higgs and a up
Higgs, and one coupling (respectively two) involves a lepton  doublet and a up Higgs.

These pairings afford a nice geometrical description. The $n=2$ locus has codimension $6$ in the invariant extension moduli space $H^1(V_2 \otimes V_3^*)_+$. The $6$ dimensional normal space
to it at an $n=2$ point has a pretty picture: it can be
identified with the space of $2 \times 3$ matrices, or linear maps $f$ from a
fixed three-dimensional space $A$ to a fixed two-dimensional space $B^*$. The
number $n$ of Higgs pairs that corresponds to such an $f$ is $n=2-{\rm rank}(f)$. The $n$-dimensional space $H^1 (\wedge^2 \tV)_-$ of up Higgs is given by the dual of
the cokernel of $f$, which is of course a subspace of $B$, the vector
space dual to $B^*$. It is two-dimensional at the origin, zero-dimensional
generically, and one-dimensional on the determinantal locus, i.e. for
those maps $f$ whose rank is $1$. The space $H^1 ( \wedge^2
\tV^*)_-$ of leptons and down Higgs consists everywhere of the two light generations of
leptons (the image of $H^1 (\wedge^2 V_2)_-$) plus the $(1+n)$-dimensional
kernel of $f$. Generically this kernel is just the heavy lepton
doublet; on the locus where $n=1$ (respectively at the origin where
$n=2$) it has dimension $2$ (respectively $3$) and there is no consistent
way to say which of these is the heavy lepton and which are the
Higgs.

One more piece of geometrical interpretation. As explained in
\cite{Bouchard:2005ag}, generically, the matrix $M^T$ has rank $17$, and we
obtain a model with no massless Higgs fields. To get $n$ Higgs pairs, the rank
of $M^T$ must decrease to $17-n$. As mentioned above, this is the case in a
codimension $2$ region for $n=1$ (and codimension $6$ for $n=2$) in the
invariant extension moduli space.
 Thus, the $42$ tangent directions for $n=1$ (the $38$
 tangent directions for
 $n=2$)\footnote{These numbers take into account
 that we quotient the invariant extension space by a
 one-dimensional space corresponding to rescaling the bundle.}
 to the special locus where we obtain $n$ Higgs pairs  should not develop
moduli-dependent Higgs $\m$-terms, exactly as we obtained in
\eqref{e:muterms}. However, if we move in the normal directions to the $n$
Higgs  pair locus, the rank of $M^T$ increases, and the pairs of Higgs
doublets should acquire a mass.

In the effective theory we have an analogous interpretation that precisely
parallels the above discussion. In the effective theory, the vector bundle
moduli fields normal to the $n$ Higgs pair locus possess the tri-linear
couplings to the Higgs fields given by \eqref{e:muterms}. When these moduli
fields  acquire non-zero vacuum expectation values (VEV's), we move to a point
in moduli space where  the effective theory is deformed in a direction
perpendicular to $n$ Higgs pair locus. In this case, these tri-linear
couplings generate non-zero $\mu$-terms for the Higgs fields, i.e. at this new
point in moduli space the Higgs field pairs become  massive.

\subsection{{\bf ($\phi$)} Triple Pairing}

The  triple pairing, involving only moduli fields, is given by
\begin{equation}
{\rm \bf (\phi)}~~~~ H^1 ( \ad \tV)_+^{(51,0)} \times H^1 ( \ad
\tV)_+^{(51,0)} \times H^1 (\ad \tV)_+^{(51,0)} \to \IC.
\end{equation}
Although we have not computed it explicitly, this pairing must be zero.
Namely,  to all orders in string perturbation the self coupling of moduli
fields is zero and thus  these fields can acquire  non-zero VEV's, which in
the effective theory parameterize deformations from a chosen locus in moduli
space. However, it is expected that non-perturbative effects introduce moduli
superpotential which would fix  VEV's of these fields; this topic is beyond
the scope of this paper.

\section{Physics Implications}\label{s:pheno}
In this section we  discuss physics  implications of the tri-linear
superpotential couplings. In order to interpret   the triple pairings
\eqref{e:tricoup} --- computed in the previous section --- as  Yukawa couplings of
the massless chiral superfields, one should always refer to table
\ref{t:lespec}, and associate particles to their respective cohomology groups.

Although the  triple pairing calculation was performed for the loci with $n=0,1,2$
massless Higgs pairs, we shall primarily focus on the interpretation of the
results for the locus with  $n=1$ massless Higgs pair.

\subsection{{\bf (u)} Triple Pairing: Up-Sector Yukawa Couplings}

For the $n=1$ Higgs pair locus, these couplings are the Yukawa couplings $\lambda^{ij}_u$ for
the up-sector quarks (the third term in the superpotential $W_1$ in \eqref{w1}). The results of the previous section reveal that  the $3\times 3$ matrix
$\lambda^{ij}_u$ is in general of rank $3$ and has the following form:
\begin{equation}
\lambda_u=\begin{pmatrix}
a & b & c\\
b & d & e\\
c & e & 0
\end{pmatrix} \label{matrix}
\end{equation}

The coefficients in the  $\lambda_u$ matrix are holomorphic functions of
moduli (tangent to the $n=1$ Higgs pair locus).  The physical Yukawa matrix
depends on the   normalization of the kinetic energy terms for the quark
fields, which can depend on Calabi-Yau and/or vector bundle moduli.
Nevertheless since  the rank of \eqref{matrix} is in general $3$, the physical Yukawa  matrix should also have rank $3$,  thus yielding non-zero masses for
all the three up-sector quarks; at special points on the $n=1$ locus one expects
to obtain a fully realistic  up-quark sector mass hierarchy.

The superpotential Yukawa couplings for the $n=2$ massless Higgs pair locus
involve two matrices of the type \eqref{matrix}, giving the couplings with the
two up Higgs fields. In this case as well, all three up-sector quark masses are
generically non-zero, and at  special points on the $n=2$ locus one expects to
obtain a fully realistic up-sector mass hierarchy.

\subsection{{\bf (d)} Triple Pairing: Down-Sector and R-parity Violating Yukawa Couplings}

The tri-linear couplings of the (${\bf d}$)-triple pairing determine the Yukawa
couplings of the down-sector quarks (the second term of $W_1$ in \eqref{w1}) and of the
charged-sector leptons (the first terms of $W_1$ in \eqref{w1}), as well as the
R-parity violating terms of $W_2$ in \eqref{w1}. As it was shown in the
previous section these couplings are {\it all} zero. The expectation is that
some of these couplings will become non-zero due to  quantum, worldsheet
instanton effects. The calculation of such effects is beyond the scope of this
paper.

Experimental constraints essentially require  the absence of  R-parity
violating couplings. Therefore,  phenomenological  viability of the model
should eventually be tested by calculations of the couplings at the
quantum level.  In an optimistic scenario, R-violating couplings could remain
zero at the quantum level for a restricted subspace of  the moduli space.

On the other hand, for the model to be  phenomenologically viable it has to
have  non-zero down-sector quark  and charged-sector lepton Yukawa couplings,
generated at the quantum level.  A non-zero Yukawa matrix of the down-sector
quarks, along with the Yukawa matrix of the up-sector quarks, also
determines the Cabibbo-Kobayashi-Maskawa (CKM) matrix, which specifies the CP
violating effects in the quark sector of the model.

\subsection{{\bf ($\mu$)} Triple Pairing: $\mu$-terms and Neutrino Yukawa Couplings}

The {\bf ($\mu$)}
        pairing corresponds to the moduli-dependent Higgs $\mu$-term, i.e. the
first term of $W_\phi$  in \eqref{wphi}, as well as  the tri-linear couplings in the neutrino sector, i.e.  the terms of $W_\nu$ in  \eqref{wnu}, where the
role of the right-handed neutrinos $\nu_{Ri}$ is played  by vector bundle
moduli $\phi_i$. This model therefore provides an interesting mechanism for
generating a $\mu$-term as well as a term responsible for giving a Dirac
mass to a neutrino.

These couplings were determined in the previous section on both the loci  with
$n=1$ and $n=2$ massless Higgs pairs. Let us first focus on the results for
the $n=1$ locus. The moduli space, perpendicular to the $n=1$ locus, is
two-dimensional.  In the effective theory it is specified by two vector bundle
moduli fields, say $\phi_1$ and $\phi_2$. These two  moduli fields have
tri-linear couplings, say, to the up-Higgs ${\bar H}$  and the down-Higgs $H$,
and to the up-Higgs ${\bar H}$ and one lepton doublet $L$, respectively:
 \begin{equation}\label{e:neut}
 W\sim \phi_1\, H\, {\bar H}\ +\  \phi_2\, L\, {\bar H}\, .
 \end{equation}
Of course, this is a specific choice for the  assignment of $H$ and $L$
fields. (Since $H$ and $L$ are in  the same representation of the SM gauge
group, their role can be interchanged, as will be explained below.)

 The  first term therefore plays the role
of a moduli-dependent Higgs $\mu$-term. Namely, after $\phi_1$ acquires a non-zero
VEV, the Higgs pair becomes massive. In order to generate an acceptable $\mu$-term, the non-zero VEV has to be proportional to the electro-weak scale (${\cal
O}$($1$ TeV)), which is much smaller than the string scale (${\cal O}$($10^{17}$ GeV)). Thus  the VEV has to be  fine-tuned, specifying a deformation in the
moduli space that is ``extremely close'' to the $n=1$ locus. Thus, we have a way
to technically obtain a phenomenologically acceptable $\mu$-term, however with
some fine-tuning.

The second term plays the role of a neutrino tri-linear coupling where  the
role of the the right-handed neutrino is played by the vector bundle modulus
$\phi_2$. After electro-weak symmetry breaking, i.e.  when ${\bar H}$
acquires a non-zero VEV, this term generates a Dirac mass for one neutrino
species.

 Note that in principle  one can  choose any other linear combination of the $\phi_1$
and $\phi_2$ fields to acquire a non-zero VEV. In this case the down-Higgs becomes a
specific combination of the $H$ and $L$ fields, and the non-zero VEV generates a
$\mu$-term for the Higgs pair. A combination of $\phi_1$ and $\phi_2$ fields,
orthogonal to the one that  acquires a non-zero VEV,  in turn corresponds to the
right-handed neutrino field. It couples to the up-Higgs and a specific
combination of the $H$ and $L$ fields that are orthogonal to the down-Higgs
field. This is now a tri-linear coupling which, after electro-weak symmetry
breaking, again generates a mass for one neutrino species.\footnote{Which
combination of $H$ and $L$ fields is interpreted as a down-Higgs field and
which one as a lepton field may be further constrained at the quantum level: the requirement that R-parity violating couplings be absent may dictate a specific combination of $H$ and $L$ fields to be a down-Higgs.}

Note also that one can remain within the $n=1$ locus where both $\phi_1$ and $\phi_2$ have zero VEV's. In this case, $\phi_1$ and $\phi_2$ can be interpreted as two right-handed neutrinos, and $(L,H)$ as two lepton doublets. Equation \eqref{e:neut} then generates masses for {\em two} neutrino species. In this case the model has no $\mu$ parameter.

The $n=2$ locus  also has a very interesting structure for these  tri-linear
couplings. As determined in the previous section there are  now six moduli
fields $\phi_{ij}$ ($i=1,2,3\, j=1,2$), transverse to the $n=2$ locus, that couple
via tri-linear couplings to the two up-Higgs fields ${\bar H }_j$ $(j=1,2)$  and
the three fields $L_i=\{ H_1,H_2,L\}$ ($i=1,2,3$). Here the  three $L_i$ fields
can be interpreted as two down-Higgs  and one lepton doublet. The tri-linear
couplings are schematically of the form:
\begin{equation}\label{e:neut2}
 W\sim \sum_{i=1}^3\sum_{j=1}^2\ \phi_{ij}\, L_i\, {\bar H}_j \, .
 \end{equation}
Note that these terms indeed provide the moduli dependent $\mu$-terms for
$n=2$ Higgs doublet pairs as well as  two candidates for the right-handed
neutrinos. For example, choosing specific non-zero VEV's for $\phi_{11}$ and
$\phi_{22}$ moduli fields generates the  two $\mu$-mass terms for both Higgs
pairs, $(H_1,\ {\bar H}_1)$ and $(H_2,\ {\bar H}_2)$, respectively.  On the
other hand, $\phi_{31}$ and $\phi_{32}$ play the role of two right-handed
neutrinos.\footnote{There are also additional SM singlet fields $\phi_{12}$
and $\phi_{21}$ that couple to the Higgs fields.} They couple to  a single lepton
doublet $L$, so after the electroweak symmetry breaking (and  assuming that
the kinetic energy terms do not have off-diagonal mixing terms for the lepton
doublet fields) there is only one massive neutrino. There is also a
possibility that  with non-zero Yukawa couplings for the charged-sector leptons
(obtained at the quantum level), the model could possess a non-trivial CKM
matrix in the lepton sector.\footnote{Of course, there is a wealth of other
moduli directions, transverse to the $n=2$ locus, that can yield $\mu$-mass
parameters for Higgs pairs, and one massive neutrino. In general, the two
down-Higgs fields and the lepton doublet will correspond to specific linear
combinations of $(H_1, \ H_2,\ L)$ fields.}

One can also remain within the $n=2$ locus where all $\phi_{i,j}$'s have zero VEV's. Now the three $L_i$ can be interpreted as lepton doublets, and the $\phi_{i,j}$'s as right-handed neutrinos. Equation \eqref{e:neut2} now generates masses for all three neutrinos.

The physics implications of the Yukawa couplings, calculated in this paper at the
classical level, are encouraging. We have demonstrated that in the  up-quark
sector one can in principle obtain a realistic mass hierarchy.  We also
demonstrated  that $\mu$ parameter(s) for the Higgs pairs(s) can be generated,
and that at least one of the neutrinos can obtain a non-zero  Dirac  mass. At the classical level, down sector quarks and charged sector leptons have zero masses. In addition, all R-parity violating terms vanish. Thus all baryon number violating processes and lepton number violating processes are absent. The
phenomenological viability of the model should  be further tested at the
quantum level where the down-sector  and charged-lepton sector Yukawa
couplings are expected to be generated. It is also at the quantum level that
the absence of R-violating couplings  is expected to impose strong
constraints on the allowed moduli subspace of the model.

\begin{ackn}
We would like to thank Volker Braun, Paul Langacker, Pran Nath
and Tony Pantev for valuable discussions. The work of R.D. is supported by an NSF grant DMS 0104354 and by an NSF Focused Research Grant DMS 0139799. The work of M.C. is supported by the DOE grant DE-FG03-95ER40917 and by the Fay R. and Eugene L. Langberg Chair. The work of V.B. is supported by an MSRI Postdoctoral Fellowship for the ``New Topological Structures in Physics" program and an NSERC Postdoctoral Fellowship. V.B. would also like to thank the Department of Mathematics of University of Pennsylvania for hospitality while part of this work was conducted.
\end{ackn}



\providecommand{\href}[2]{#2}\begingroup\raggedright\endgroup


\begin{thebibliography}{10}

\bibitem{Bouchard:2005ag}
V.~Bouchard and R.~Donagi, {\it An SU(5) heterotic standard model},  {\em Phys.
  Lett.} {\bf B633} (2006) 783--791,
  [\href{http://xxx.lanl.gov/abs/hep-th/0512149}{{\tt hep-th/0512149}}].

\bibitem{Donagi:2004ub}
R.~Donagi, Y.-H. {H}e, B.~A. {O}vrut, and R.~{R}einbacher, {\it The spectra of
  heterotic standard model vacua},  {\em {JHEP}} {\bf 06} (2005) 070,
  [\href{http://xxx.lanl.gov/abs/hep-th/0411156}{{\tt hep-th/0411156}}].

\bibitem{CleaverBuchmuller}
G.~B. Cleaver, A.~E. Faraggi, and D.~V. Nanopoulos, {\it A minimal superstring
  standard model. I: Flat directions},  {\em Int. J. Mod. Phys.} {\bf A16}
  (2001) 425--482, [\href{http://xxx.lanl.gov/abs/hep-ph/9904301}{{\tt
  hep-ph/9904301}}]; W.~Buchmuller, K.~Hamaguchi, O.~Lebedev, and M.~Ratz, {\it The supersymmetric
  standard model from the heterotic string},
  \href{http://xxx.lanl.gov/abs/hep-ph/0511035}{{\tt hep-ph/0511035}}.

\bibitem{CSUrev}
M.~Cveti{\v c}, G.~Shiu, and A.~M. Uranga, {\it Chiral four-dimensional N = 1
  supersymmetric type IIA orientifolds from intersecting D6-branes},  {\em
  Nucl. Phys.} {\bf B615} (2001) 3--32,
  [\href{http://xxx.lanl.gov/abs/hep-th/0107166}{{\tt hep-th/0107166}}];
R.~Blumenhagen, M.~Cveti{\v c}, P.~Langacker, and G.~Shiu, {\it Toward
  realistic intersecting D-brane models},
  \href{http://xxx.lanl.gov/abs/hep-th/0502005}{{\tt hep-th/0502005}}.

\bibitem{Dijkstra:2004cc}
T.~P.~T. Dijkstra, L.~R. Huiszoon, and A.~N. Schellekens, {\it Supersymmetric
  standard model spectra from RCFT orientifolds},  {\em Nucl. Phys.} {\bf B710}
  (2005) 3--57, [\href{http://xxx.lanl.gov/abs/hep-th/0411129}{{\tt
  hep-th/0411129}}].

\bibitem{sharpe}
  S.~Katz and E.~Sharpe,
  {\it Notes on certain (0,2) correlation functions},
  Commun.\ Math.\ Phys.\  {\bf 262}, 611 (2006)
  [\href{http://xxx.lanl.gov/abs/hep-th/0406226}{{\tt hep-th/0406226}}]; 
E.~Sharpe, {\it Notes on correlation functions in (0,2) theories},
  \href{http://xxx.lanl.gov/abs/hep-th/0502064}{{\tt hep-th/0502064}}.

\bibitem{CPetal}
M.~Cveti{\v c} and I.~Papadimitriou, {\it Conformal field theory couplings for
  intersecting D-branes on orientifolds},  {\em Phys. Rev.} {\bf D68} (2003)
  046001, [\href{http://xxx.lanl.gov/abs/hep-th/0303083}{{\tt
  hep-th/0303083}}]; 
D.~Cremades, L.~E. Ib{\' a}{\~ n}ez, and F.~Marchesano, {\it Yukawa couplings
  in intersecting D-brane models},  {\em JHEP} {\bf 07} (2003) 038,
  [\href{http://xxx.lanl.gov/abs/hep-th/0302105}{{\tt hep-th/0302105}}]; 
D.~L{\" u}st, P.~Mayr, R.~Richter, and S.~Stieberger, {\it Scattering of gauge,
  matter, and moduli fields from intersecting branes},  {\em Nucl. Phys.} {\bf
  B696} (2004) 205--250, [\href{http://xxx.lanl.gov/abs/hep-th/0404134}{{\tt
  hep-th/0404134}}]; M.~Bertolini, M.~Billo, A.~Lerda, J.~F. Morales, and 
R.~Russo, {\it Brane world
  effective actions for D-branes with fluxes},
  \href{http://xxx.lanl.gov/abs/hep-th/0512067}{{\tt hep-th/0512067}}.

\bibitem{Braun:2005nv}
V.~Braun, Y.-H. He, B.~A. Ovrut, and T.~Pantev, {\it The exact MSSM spectrum
  from string theory},  \href{http://xxx.lanl.gov/abs/hep-th/0512177}{{\tt
  hep-th/0512177}}.

\bibitem{Braun:2006ae}
V.~Braun, Y.-H. He, and B.~A. Ovrut, {\it Stability of the minimal heterotic
  standard model bundle},  \href{http://xxx.lanl.gov/abs/hep-th/0602073}{{\tt
  hep-th/0602073}}.

\bibitem{Gomez:2005ii}
T.~L. Gomez, S.~Lukic, and I.~Sols, {\it Constraining the K\"ahler moduli in the
  heterotic standard model},
  \href{http://xxx.lanl.gov/abs/hep-th/0512205}{{\tt hep-th/0512205}}.

\bibitem{Braun:2005ux}
V.~Braun, Y.-H. {H}e, B.~A. {O}vrut, and T.~{P}antev, {\it A heterotic standard
  model},  \href{http://xxx.lanl.gov/abs/hep-th/0501070}{{\tt hep-th/0501070}}.

\bibitem{Nath:2006ut}
P.~Nath and P.~F. Perez, {\it Proton stability in grand unified theories, in
  strings, and in branes},  \href{http://xxx.lanl.gov/abs/hep-ph/0601023}{{\tt
  hep-ph/0601023}}.

\bibitem{Donagi:2000si}
R.~{D}onagi, {B}.~ {O}vrut, {T}.~ {P}antev, and {D}.~ {W}aldram, {\it
  Spectral involutions on rational elliptic surfaces},  {\em Advances in
  {T}heoretical and {M}athematical {P}hysics} {\bf 5} (2002) 499,
  [\href{http://xxx.lanl.gov/abs/math.{AG}/0008011}{{\tt math.{AG}/0008011}}].

\bibitem{Donagi:2000sm}
R.~{D}onagi, {B}.~ {O}vrut, {T}.~ {P}antev, and {D}.~ {W}aldram, {\it
  Standard-model bundles},  {\em Advances in {T}heoretical and {M}athematical
  {P}hysics} {\bf 5} (2002) 563,
  [\href{http://xxx.lanl.gov/abs/math.{AG}/0008010}{{\tt math.{AG}/0008010}}].

\bibitem{Donagi:2000zf}
R.~Donagi, B.~A. Ovrut, T.~Pantev, and D.~Waldram, {\it Standard-model bundles
  on non-simply connected Calabi-Yau threefolds},  {\em JHEP} {\bf 08} (2001)
  053, [\href{http://xxx.lanl.gov/abs/hep-th/0008008}{{\tt hep-th/0008008}}].

\bibitem{GKLN}
J.~Giedt, G.~L. Kane, P.~Langacker, and B.~D. Nelson, {\it Massive neutrinos
  and (heterotic) string theory},  {\em Phys. Rev.} {\bf D71} (2005) 115013,
  [\href{http://xxx.lanl.gov/abs/hep-th/0502032}{{\tt hep-th/0502032}}].

\bibitem{Green:1987mn}
M.~B. Green, J.~H. {S}chwarz, and E.~{W}itten, {\em Superstring {T}heory.
  {V}ol. 2: {L}oop {A}mplitudes, {A}nomalies and {P}henomenology}.
\newblock Cambridge {M}onographs {O}n {M}athematical {P}hysics. Cambridge
  {U}niversity {P}ress, Cambridge, {UK}, 1987.
\newblock 596 p.

\end{thebibliography}

\end{document}